\documentstyle[twoside,fleqn,espcrc2,epsf]{article}

\newcommand{\pseudo}{{P}}
\newcommand{\err}[2]{{\footnotesize {$\;\begin{array}{@{}l@{}}
			  +\makebox[1.3em][r]{#1} \\[-0.4em]
			  -\makebox[1.3em][r]{#2}
			\end{array}$}}}

\newcommand{\mev}{{\rm MeV}}

\newcommand{\AmS}{{\protect\the\textfont2
  A\kern-.1667em\lower.5ex\hbox{M}\kern-.125emS}}

\title{Charmed Meson Spectroscopy and Matrix Elements with an
$O(a)$-Improved Clover Fermion Action}

\author{{\it UKQCD Collaboration}\\
presented by David G.~Richards\\[0.25cm]
Department of Physics, University of
Edinburgh \\ Mayfield Road, Edinburgh EH9 3JZ, Scotland}

\begin{document}

\begin{abstract}
We present preliminary results for the spectrum and decay matrix
elements for heavy-light and heavy-heavy mesons, obtained on the
64-node Meiko Computing Surface at the University of Edinburgh.

Quark propagators are computed with an $O(a)$-improved fermion action
on $24^3 \times 48$ lattices at $\beta = 6.2$, using three values of
the quark mass up to around the strange quark mass, and four values of
the quark mass in the region of the charm quark mass.  We compare
results for the hyperfine splitting in charmonium with those obtained
using the conventional Wilson fermion action and find that the
splitting is 1.83(15) times larger with the improved action.  Our
measurements of $f_B$ indicate non-scaling corrections of the order of
20\% to the Heavy Quark Effective Theory expectation.  A comparison is
made with results obtained on $16^3
\times 48$ lattices at $\beta = 6.0$.
\end{abstract}

\maketitle

\section{INTRODUCTION}

This contribution represents the talk ``Latest Results from the UKQCD
Collaboration'' presented at Lattice 92.

As the masses of hadrons in lattice gauge simulations approach the
inverse lattice spacing, $a^{-1}$, quantities are expected to become
increasingly sensitive to discretisation errors.  Thus it is in this
regime, corresponding in present simulations to quarks with masses of
the charm and above, that we expect differences between an
$O(a)$-improved fermion action and the standard Wilson fermion action
to be most apparent.

In this talk I shall describe the recent studies by the UKQCD
Collaboration of mesons containing propagating charm quarks.  The talk
begins with a brief description of the computational details.  Then a
comparison is made of the hyperfine splitting in charmonium between
the standard Wilson action, and the $O(a)$-improved clover
action~\cite{clover}.  For the clover action, this investigation is
extended to the $D$ and $B$ systems.  Finally, measurements are made
of the pseudoscalar decay constant, $f_P$, for heavy-light systems.
By determining the behaviour of $f_P
\surd M_{\pseudo}$ as a function of the pseudoscalar mass,
$M_{\pseudo}$, an estimate is made of the magnitude of the non-scaling
terms in the Heavy Quark Effective Theory (HQET) in the region of the
$B$-meson mass.

\section{COMPUTATIONAL DETAILS}

Clover quark propagators were generated on quenched $24^3 \times 48$
lattices at $\beta = 6.2$, using three values of the quark mass up to
the region of the strange quark mass, with hopping parameters
$\kappa_l = 0.14144, 0.14226
\mbox{ and } 0.14262$~\cite{ukqcd92}.
In addition, propagators were generated at
four values of the quark mass in the region of the charm quark mass,
with hopping parameters $\kappa_h = 0.133, 0.129, 0.125 \mbox{ and }
0.121$; the charm quark mass corresponds to $\kappa_h \simeq 0.129$.
The preliminary results presented in this write-up are based on an
analysis of 33 configurations, the results in the talk being based on
an analysis of 21 configurations.

The light quark propagators were generated using local sources and
sinks (LL).  The charm quark propagators were computed from smeared
sources, to both local sinks (LS) and smeared sinks (SS), using a
Jacobi-smearing algorithm~\cite{sara92} with a smearing radius $r
\simeq 5$.

The comparison of the hyperfine splitting in charmonium between the
Wilson and clover fermion actions was obtained on a common sample of
18 configurations, using local sources and sinks~\cite{hyper92}.
All fits to the correlators are performed taking into account the
correlations between different time slices, with the error bars
obtained using a bootstrap procedure.

\section{HYPERFINE SPLITTING}

Quenched calculations with the standard Wilson fermion action
typically find values of vector-pseudoscalar mass splittings,
$m_{D^\ast} - m_D$ and $m_{J/\psi} - m_{\eta_c}$, which are much too
small, indeed typically by a factor of 2 and 4 respectively at $\beta
= 6.2$~\cite{bochicchio92,abada92}.  The clover fermion action is
related to the standard Wilson fermion action through the addition of
a term,
\begin{equation}
S_F^C = S_F^W - i \frac{\kappa}{2}
\sum_{x, \mu, \nu} \bar{q}(x)
F_{\mu \nu}(x) \sigma_{\mu \nu} q(x). \label{eq:clover_term}
\end{equation}
The hyperfine splitting is expected
to be sensitive to the magnetic moment form of this additional term.

\begin{table*}[hbt]
\setlength{\tabcolsep}{1.5pc}
\newlength{\digitwidth} \settowidth{\digitwidth}{\rm 0}
\catcode`?=\active \def?{\kern\digitwidth}
\caption{Masses (in lattice units) of the pseudoscalar and vector
heavy-heavy mesons.}
\label{tab:masses}
\begin{tabular} {crrr} \hline
 & $m_{\eta_c}a$ & $m_{J/\psi}a$ & $(m_{J/\psi}-m_{\eta_c})a$ \\
 \hline
Wilson & 1.066\err{6}{4} & 1.076\err{4}{5} & 0.0104\err{8}{7} \\
clover & 1.071\err{6}{4} & 1.088\err{5}{5} & 0.0190\err{12}{16} \\ \hline
\end{tabular}
\end{table*}

To compare the values for the hyperfine splitting in charmonium, we
chose $\kappa = 0.135$ for the Wilson action, and $\kappa = 0.129$ for
the clover action, yielding approximately the same value of the
pseudoscalar mass, $m_{\eta_c}$, for the two actions.  The measured
values of the vector and pseudoscalar masses are shown in
Table~\ref{tab:masses}.  Setting the scale from the string tension
gives $a^{-1} = 2.73(5)$ GeV~\cite{ukqcd92}, and we find the values of
the masses quoted in the table to be within a few percent of their
physical values.  However, the mass splitting quoted in the third
column is very different for the two actions, and we find that
\begin{equation}
\frac{(m_{J/\psi}-m_{\eta_c})^{\rm clover}}
                              {(m_{J/\psi}-m_{\eta_c})^{\rm Wilson}}
                                                 =1.83\mbox{\err{13}{15}}
\label{eq:ratio}
\end{equation}
Using the value of $a^{-1}$ quoted above, we obtain
\begin{eqnarray}
m_{J/\psi}-m_{\eta_c} & = & 28\mbox{\err{2}{2}}\ \mev \mbox{ ~~(Wilson)}
\label{eq:hfsw} \nonumber \\
m_{J/\psi}-m_{\eta_c} & = & 52\mbox{\err{3}{4}}\  \mev \mbox{ ~~(clover).}
\label{eq:hfsc}
\end{eqnarray}
to be compared to the experimental value of 117(2)~MeV.  The
corresponding values found by El-Khadra et al. \cite{fnal91}, using a
mean field calculation of the coefficient of the clover term in
eq.~\ref{eq:clover_term}, are: 51(3) MeV at $\beta = 5.7$, 62(4) MeV
at $\beta = 5.9$ and 68(5) MeV at $\beta = 6.1$, where the value of
the string tension is obtained from the 1P-1S mass splitting.
Extrapolating linearly in $a^2$ to the continuum limit, they obtain
\begin{equation}
m_{J/\psi}-m_{\eta_c} = 73 \pm 10 \mev.
\end{equation}
This is still far short of the experimental value.  However, it has
been argued that the remaining discrepancy is consistent with the
corrections arising from the quenched approximation~\cite{aida92}.

\begin{figure}[htb]
\centerline{
	\epsfysize=150pt
	\epsfbox[00 80 580 550]{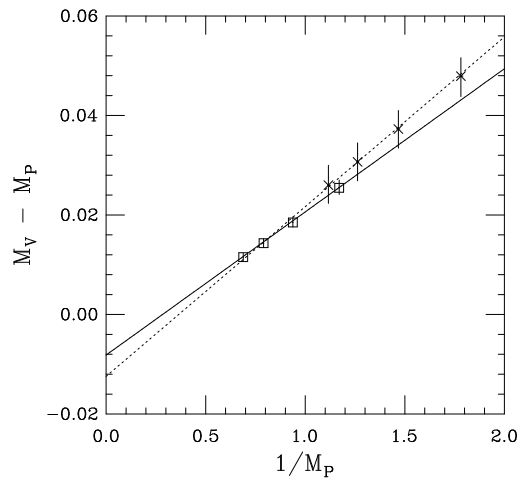}
	}
\caption{The hyperfine mass splitting $M_V - M_\pseudo$ against
$1/M_\pseudo$, in lattice units, is shown both for heavy-heavy mesons
(squares), and for heavy-light mesons after extrapolation of
$\kappa_l$ to $\kappa_{\rm crit} = 0.1431$ (crosses).  The fits to the
heavy-heavy and heavy-light data are indicated by the solid and dashed
lines respectively.}
\label{fig:mv_minus_mp}
\end{figure}
Figure~\ref{fig:mv_minus_mp} shows $M_V - M_\pseudo$ (where $V$ and
$\pseudo$ represent the vector and pseudoscalar respectively) for both
heavy-light and heavy-heavy mesons obtained using the clover action on
the 33 configurations.  Fits are over the range 10 to
16, using the LS charm propagators.  For both the heavy-light and
heavy-heavy systems, linear behaviour with $1/M_\pseudo$ is apparent;
the non-zero intercept of the fits is presumably a manifestation of
lattice artifacts.  Setting $a^{-1} = 2.73\mbox{ GeV}$, we obtain
%
\begin{eqnarray}
m_{D^{\ast}} - m_{D} & = & 102\mbox{\err{10}{9}}\ \mev \nonumber \\
m_{B^{\ast}} - m_{B} & = & 14\mbox{\err{14}{10}}\ \mev,
\end{eqnarray}
to be compared to the experimental values of 146(1)~MeV and 46(3)~MeV
respectively.  An analysis using the clover action at $\beta = 6.0$,
for $\kappa_l$ in the region of the strange quark mass, yields
$m_{D^{\ast}} - m_{D} = 76\mbox{\err{17}{21}}$~\cite{dsh92},
clearly indicating a remaining sensitivity to the lattice spacing,
$a$.

\section{DECAY CONSTANT $f_P$}

\begin{figure}[htb]
\centerline{
	\epsfysize=150pt
	\epsfbox[00 80 580 550]{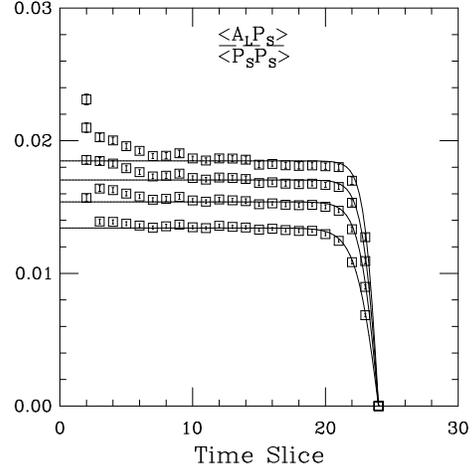}
	}
\caption{The ratio in eq.~6
is shown for all four
$\kappa_h$, at fixed $\kappa_l = 0.14226$.  The solid lines represent the
fits to the data.}
\label{fig:AP_over_PP}
\end{figure}

 We extract the heavy-light pseudoscalar decay constant $f_P$ through
a fit to the ratio,
\begin{equation}
\frac{\displaystyle \sum_{\vec{x}} \langle A^4_L(\vec{x},t) P^{\dagger}_S(0)
\rangle}
{\displaystyle \sum_{\vec{x}} \langle P_S(\vec{x},t) P^{\dagger}_S(0)
\rangle} \sim \frac{f_P}{Z_S} \tanh{M_\pseudo ( T/2 - t)}
\label{eq:AP_over_PP}
\end{equation}
where $A_\mu$ is the axial vector current, $P = \bar{\psi} \gamma_5
\psi$, and $Z$ is defined through
\begin{equation}
Z = \langle 0 | P | P \rangle.
\end{equation}
$S$ and $L$ refer to quantities constructed from smeared and local
quark propagators respectively.  $Z_S$ and $M_\pseudo$ are determined
from the fit to $\langle P_S(\vec{x},t) P^{\dagger}_S(0)
\rangle$.
The fits both to eq.~\ref{eq:AP_over_PP} and to the pseudoscalar
correlator are over the range 10~-~16.  The quality of the data for
eq.~\ref{eq:AP_over_PP}, together with the fits to the data, are
illustrated in Figure~\ref{fig:AP_over_PP}.

\begin{figure}[htb]
\centerline{
	\epsfysize=150pt
	\epsfbox[00 80 580 550]{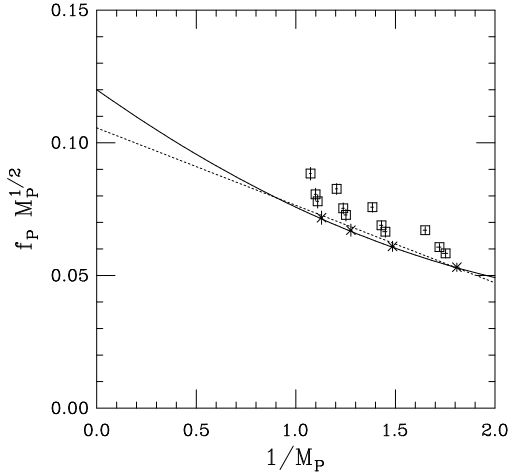}
	}
\caption{
$f_P \surd M_P$ against $1/M_P$ in lattice units is shown for all
$\kappa_l$ and $\kappa_h$ (squares).  Also shown are the values after
linear extrapolation of $\kappa_l$ to $\kappa_{\rm crit} = 0.1431$
(crosses).  The solid and dotted lines represent a quadratic fit to
all four points, and a linear fit to the heaviest three points
respectively, at $\kappa_{\rm crit}$.}
\label{fig:fP_all_kappas}
\end{figure}

The expected behaviour from the HQET is
\begin{equation}
f_P \surd M_\pseudo \sim A + O(1/M_\pseudo)
\end{equation}
where A is a constant, up to logarithmic terms.
Figure~\ref{fig:fP_all_kappas} shows $f_P \surd M_P$ against $M_P$ for
all $\kappa_l$ and $\kappa_h$, together with the value after the
extrapolation of $\kappa_l$ to $\kappa_{\rm crit} = 0.1431$~\cite{ukqcd92}.

We determine the behaviour of $f_P \surd M_P$ with $1/M_\pseudo$, at
$\kappa_{\rm crit}$, from a correlated fit to
\begin{equation}
f_P \surd M_P = a_1 + \frac{a_2}{M_P} + \frac{a_3}{M_P^2}.
\label{eq:fP_form}
\end{equation}
In lattice units, a linear fit to the three heaviest points yields,
\begin{eqnarray}
a_1 & = & 0.106\mbox{\err{6}{5}} \nonumber \\
a_2 & = & -0.029\mbox{\err{2}{3}},
\label{eq:fP_lin_form}
\end{eqnarray}
with $\chi^2/\mbox{dof} = 2.4/1$,
while a quadratic fit to all four points yields
\begin{eqnarray}
a_1 & = & 0.120\mbox{\err{9}{8}} \nonumber \\
a_2 & = & -0.053\mbox{\err{8}{9}} \nonumber \\
a_3 & = & 0.018\mbox{\err{5}{4}}.
\label{eq:fP_quad_form}
\end{eqnarray}
with  $\chi^2/\mbox{dof} = 0.1/1$.
A linear fit to all four points produces a noticeably poor fit, with
$\chi^2/\mbox{dof} = 12/2$.

Using the linear fit to the heaviest three points, we obtain
\begin{eqnarray}
f_D & = & 0.204\mbox{\err{6}{6}} \mbox{~~~~~}
(\frac{a^{-1}}{2.73}) \mbox{~~~ GeV}
\nonumber \\
f_B & = & 0.174\mbox{\err{8}{7}} \mbox{~~~~~}
(\frac{a^{-1}}{2.73}) \mbox{~~~ GeV,}
\end{eqnarray}
while from the quadratic fit we obtain
\begin{eqnarray}
f_D & = & 0.198\mbox{\err{5}{5}} \mbox{~~~~~}
(\frac{a^{-1}}{2.73}) \mbox{~~~ GeV}
\nonumber \\
f_B & = & 0.183\mbox{\err{10}{8}} \mbox{~~~~~}
(\frac{a^{-1}}{2.73}) \mbox{~~~ GeV,}
\end{eqnarray}
where we take $Z_A = 1 - 0.02 g^2$, with $g$ the bare coupling.  The
renormalisation coefficient, $Z_A$, for the clover action is much
closer to unity than that for the Wilson action ($Z_A^W \simeq 1 -
0.132 g^2$).  Thus the uncertainty due to the choice of expansion
parameter~\cite{lepage91}, $g$, is only about 1\% for the clover
action, compared to about 10\% for the standard Wilson action.

The determination of $f_B$ is clearly more sensitive to the form of
the extrapolation than that of $f_D$.  An analysis at $\beta =
6.0$~\cite{jns92}, using a hopping parameter expansion of the heavy
quark propagator, favours a quadratic fit, yielding similar
coefficients (in physical units) to those quoted in
eq.~\ref{eq:fP_quad_form}. The non-scaling corrections to $f_B$ are
approximately 15\% and 20\% of the scaling expectation, obtained from
the constant term $a_1$, for the linear and quadratic fits
respectively.

\section{CONCLUSIONS}

The improvement in the measured value of the hyperfine splitting both
for heavy-heavy and heavy-light mesons using an $O(a)$-improved action
is very encouraging, and the remaining discrepancies with experiment
may be due to the effects of quenching.

The preliminary results for $f_D$ and $f_B$ presented here need
further refinement, in particular to take account of the uncertainties
in the lattice spacing.  Nevertheless, they indicate that non-scaling
contributions to the HQET predictions are substantial, even at the
mass of the $B$ meson.  It is particularly important to measure the
static results, with light quarks computed using the clover action, on
the same set of configurations.  A consistent extrapolation between
the propagating quark results presented here and the static
predictions would provide powerful evidence for the reliability of
lattice studies of heavy quarks.

\subsection*{Acknowledgements}

This research is supported by the UK Science and Engineering Research
Council under grants GR/G~32779, GR/H~49191 \& GR/H~01069, by the
University of Edinburgh and by Meiko Ltd..

\end{document}